\documentclass[prl,aps,singlecolumn,superscriptaddress]{revtex4}
\usepackage{amssymb,amsfonts,amsmath}
\usepackage{epsfig}
\usepackage[section]{placeins}

\begin{document}

\title{Spin currents in a coherent exciton gas}
\author{A.A.~High}
\affiliation{Department of Physics, University of California at San Diego, La Jolla, CA 92093-0319, USA}
\author{A.T.~Hammack}
\affiliation{Department of Physics, University of California at San Diego, La Jolla, CA 92093-0319, USA}
\author{J.R.~Leonard}
\affiliation{Department of Physics, University of California at San Diego, La Jolla, CA 92093-0319, USA}
\author{Sen Yang}
\affiliation{Department of Physics, University of California at San Diego, La Jolla, CA 92093-0319, USA}
\author{L.V.~Butov}
\affiliation{Department of Physics, University of California at San Diego, La Jolla, CA 92093-0319, USA}
\author{T.~Ostatnick{\' y}}
\affiliation{Faculty of Mathematics and Physics, Charles University in Prague, Ke Karlovu 3, 121 16 Prague, Czech Republic}
\author{M.~Vladimirova}
\affiliation{Universit{\'e} Montpellier 2, CNRS, Laboratoire Charles Coulomb UMR 5221, F-34095, Montpellier, France}
\author{A.V.~Kavokin}
\affiliation{Universit{\'e} Montpellier 2, CNRS, Laboratoire Charles Coulomb UMR 5221, F-34095, Montpellier, France}
\affiliation{School of Physics and Astronomy, University of Southampton, SO17 1BJ, Southampton, United Kingdom}
\affiliation{Spin Optics Laboratory, State University of Saint Petersburg, 1, Ulianovskaya, 198504, Russia}
\author{T.C.H.~Liew}
\affiliation{Mediterranean Institute of Fundamental Physics, 31, via Appia Nuova, Rome, 00040, Italy}
\author{K.L.~Campman}
\affiliation{Materials Department, University of California at Santa Barbara, Santa Barbara, CA 93106-5050, USA}
\author{A.C.~Gossard}
\affiliation{Materials Department, University of California at Santa Barbara, Santa Barbara, CA 93106-5050, USA}
\date{\today}

\begin{abstract}
\noindent {Spin currents and spin textures are observed in a coherent gas of indirect excitons. Applied magnetic fields bend the spin current trajectories and transform patterns of linear polarization from helical to spiral and patterns of circular polarization from four-leaf to bell-like-with-inversion.}
\end{abstract}

\maketitle

\begin{figure*}
\includegraphics[width=17cm]{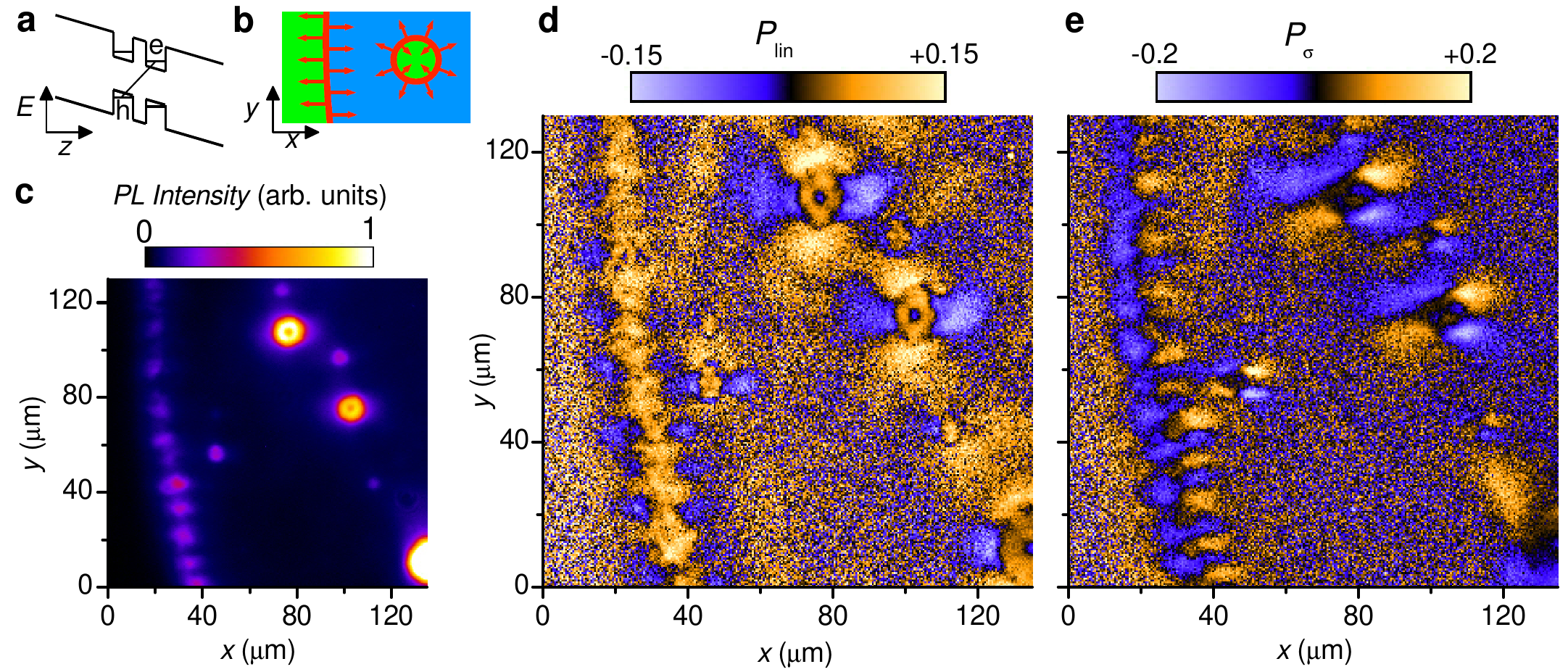}
\caption{{\bf Polarization patterns in exciton emission.} (a) Diagram of the CQW; e, electron; h, hole. (b) Schematic of exciton formation in the external ring (left) and LBS ring (right); Excitons (red) form on the boundary of hole-rich (blue) and electron-rich (green) areas. Exciton transport is indicated by red arrows. (c) A segment of the emission pattern of indirect excitons showing the external ring (left) and multiple LBS. (d,e) Patterns of linear $P_{lin} = (I_x - I_y) / (I_x + I_y)$ (d) and circular $P_{\protect\sigma} = (I_{\protect\sigma^+} - I_{\protect\sigma^-}) / (I_{\protect\sigma^+} + I_{\protect\sigma^-})$ (e) polarization of the emission of indirect excitons in the region shown in (c). $T_{bath} = 0.1$ K.}
\end{figure*}

Studies of electron spin currents in semiconductors led to the discoveries of the spin Hall effect \cite{Dâyakonov71, Hirsch99, Murakami03, Sinova04, Kato04, Wunderlic05}, persistent spin helix \cite{Koralek09}, and spin drift, diffusion, and drag \cite{Kikkawa99, Weber05, Crooker05}. There is also considerable interest in developing semiconductor (opto)electronic devices based on spin currents. An important role in spin current phenomena is played by spin-orbit (SO) coupling. It is the origin of the spin Hall effect and persistent spin helix. It also creates spin structures with the spin vector perpendicular to the momentum of the electrons in metals \cite{Hoesch04} and topological insulators \cite{Koning08, Moore10, Hasan10}. While phenomena caused by SO coupling are ubiquitous in fermionic systems, they have yet to be explored in bosonic matter. Available experimental data for bosons include the optical spin Hall effect in photonic systems \cite{Kavokin05, Leyder07, Maragkou11} and spin patterns in atomic condensates \cite{Sadler06, Lin11}. Here, we report the observation of spin currents and associated rich variety of polarization patterns in a coherent gas of indirect excitons. Applied magnetic fields bend the spin current trajectories and transform patterns of linear polarization from helical to spiral and patterns of circular polarization from four-leaf to bell-like-with-inversion. We also present a theory of exciton transport with spin precession that reproduces the observed exciton polarization patterns and indicates trajectories of spin currents.

Excitons - bound pairs of electrons and holes - form a model system to study spin currents of bosons \cite{Keldysh68}. SO coupling for an exciton originates from the combined Dresselhaus and Rashba effects for the electron and the hole \cite{Rashba88, Wu08, Luo10}. An indirect exciton can be formed by an electron and a hole confined in separate quantum-well (QW) layers (Fig. 1a). The spatial separation reduces the overlap of electron and hole wavefunctions thus producing indirect excitons with lifetimes orders of magnitude longer than those of direct excitons \cite{Lozovik76, Fukuzawa90}. Due to their long lifetimes, indirect excitons can travel over large distances before recombination \cite{Hagn95} and can cool down below the temperature of quantum degeneracy and form a condensate \cite{High12}.

The condensation of indirect excitons was predicted to cause the suppression of exciton scattering \cite{Lozovik76}. The measured strong enhancement of the exciton coherence length \cite{High12} experimentally shows the suppression of exciton scattering. The suppression of scattering results in the suppression of the Dyakonov-Perel and Elliott-Yafet mechanisms of spin relaxation \cite{Dyakonov08}. Furthermore, the spatial separation between an electron and a hole suppresses the Bir-Aronov-Pikus mechanism of spin relaxation for indirect excitons \cite{Maialle93, Leonard09}. The suppression of these mechanisms of spin relaxation results in a strong enhancement of the spin relaxation time in a condensate of indirect excitons. While the spin relaxation times of free electrons and holes can be short \cite{Maialle93}, the formation of a coherent gas of their bosonic pairs results in a strong enhancement of their spin relaxation times, facilitating long-range spin currents.

Previous studies identified the external ring and localized bright spot (LBS) rings in the emission pattern of indirect excitons (Fig. 1c) as sources of cold excitons \cite{High12}. These rings form on the boundaries between electron-rich and hole-rich regions; the former is created by current through the structure (specifically, by the current filament at the LBS center in the case of the LBS ring), whereas the latter is created by optical excitation (Fig. 1b), see \cite{High12} and references therein. In \cite{High12}, we presented the studies of spontaneous coherence of indirect excitons in the ring region.

Here, we present the studies of spin currents and associated spin patterns around the sources of cold excitons - the rings. We present the observation of patterns of circular polarization, corresponding to spin perpendicular to the QW plane. We also show that the observed polarization patterns are controlled by magnetic field: these data prove that the pattern of linear polarization corresponds to spin orientation rather than merely to the orientation of an exciton dipole. We also deduce trajectories of electron and hole spin currents from the measured polarization patterns.

The exciton polarization currents and associated spin textures are revealed by the polarization pattern of the emitted light measured by polarization-resolved imaging (Fig. 1d,e). Experiments are performed in an optical dilution refrigerator. The photoexcitation is nonresonant and spatially separated so that the exciton polarization is not induced by the pumping light.

\begin{figure*}[h]
\includegraphics[width=17cm]{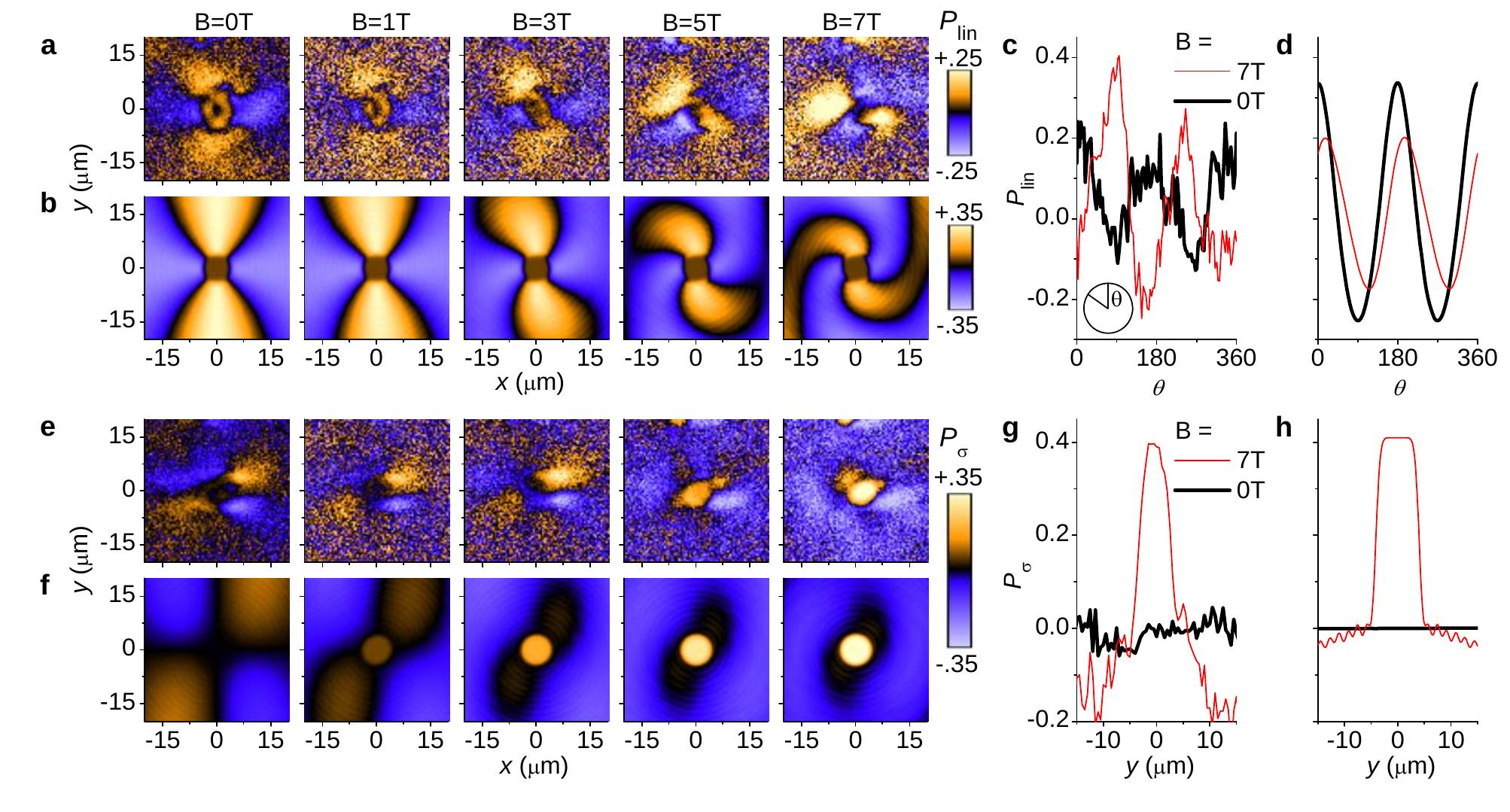}
\caption{{\bf Control of polarization patterns.} Measured (a) and simulated (b) patterns of linear polarization of the emission of indirect excitons $P_{lin}$ in the region of LBS for different magnetic fields perpendicular to the QW plane $B$. Measured (c) and simulated (d) azimuthal variation of $P_{lin}$ at a distance from the LBS center $r = 9$ $\protect\mu$m for $B = 0$ (black) and 7 T (red). Angles are measured from the $y$-axis. Measured (e) and simulated (f) patterns of circular polarization of the emission of indirect excitons $P_{\protect\sigma}$ in the region of LBS for different $B$. Measured (g) and simulated (h) cross sections of $P_{\protect\sigma}$ at $x = 0$ for $B = 0$ (black) and 7 T (red). For all data, $T_{bath} = 0.1$ K, the LBS is at (105,75) in Fig.~1c. See \cite{SI}.}
\end{figure*}

The binding energy released at the exciton formation in the rings and the current filament at the LBS center heat the exciton gas. The former heating source depletes the exciton condensate in the rings \cite{High12}. The latter is so strong that no condensate forms in the LBS ring center and the exciton gas is classical there \cite{High12}. Excitons cool down with increasing distance $r$ away from the heating sources so that they can approach the condensation temperature at $r = r_0$ where the condensation is detected by interferometric measurements \cite{High12}.

The indirect excitons in GaAs CQW may have four spin projections on the $z$ direction normal to the CQW plane: $J_z = -2, -1, +1, +2$. The states $J_z = -1$ and $+1$ contribute to left- and right-circularly polarized emission and their coherent superposition to linearly polarized emission, whereas the states $J_z = -2$ and $+2$ are dark \cite{Maialle93}. The electron and hole spin projections on the $z$-axis are given by $J_z$, while in-plane projections of electron and hole spins can be deduced from the off-diagonal elements of the exciton spin density matrix, which can be obtained from the measured polarization pattern \cite{SI}. The exciton states linearly polarized along the axes of symmetry are generally split due to in-plane anisotropy induced by the crystallographic axis orientation and strain.

The observed polarization patterns are qualitatively similar for both sources of cold excitons -- the external ring and LBS ring. An LBS ring is close to a model radially symmetric source of excitons with a divergent (hedgehog) momentum distribution (Fig. 1b) and we concentrate on the polarization textures around the LBS here. All LBS rings in the emission pattern show similar spin textures (Fig. 1d,e).

A ring of linear polarization is observed around each LBS center (Fig. 1d, 2a). This ring is observed in the region $r < r_0$ where the exciton gas is classical. The linear polarization originates from the thermal distribution of excitons over the linearly polarized exciton states. Heating of the exciton gas by the current filament reduces the polarization degree in the LBS center and, as a result, leads to the appearance of a ring of linear polarization. No such polarization reduction is observed in the external ring, consistent with the absence of heating by current filaments in the external ring area (Fig. 1d).

A helical exciton polarization texture that winds by $2\pi$ around the origin emerges in the LBS area at $r > r_0$ where the condensate forms (the latter is measured by shift-interferometry), Fig. 1d, 2a, 2c. The LBS exhibits a divergent hedgehog-shaped momentum distribution (Fig. 3d). The exciton polarization is perpendicular to exciton momentum (Fig. 2a, 3d). This produces vortices of linear polarization which emerge in concert with spontaneous coherence below the critical temperature \cite{SI}. The observed radial exciton polarization currents are associated with spin currents carried by electrons and holes bound into excitons as detailed below.

Applied magnetic fields bend the spin current trajectories creating spiral patterns of linear polarization around the origin (Fig. 2a,c). The spiral direction of the exciton polarization current clearly deviates from the radial direction of the exciton density current (Fig. 2a,c). The control of the polarization patterns by magnetic field shows that they are associated with spin.

Regular patterns are also observed in circular polarization (Fig. 1e, 2e). An LBS source of excitons generates a four-leaf pattern of circular polarization (Fig. 1e, 2e). This pattern vanishes with increasing temperature \cite{SI}. An applied magnetic field transforms the four-leaf pattern to a bell-like pattern of circular polarization with a strong circular polarization in the center and polarization inversion a few $\mu$m away from the center (Fig. 2e,g).

Polarization patterns are also observed in the external ring region (Fig. 1d,e). At low temperatures, the macroscopically ordered exciton state (MOES) forms in the external ring. The MOES is characterized by a spatially ordered array of higher-density beads and is a condensate in momentum space \cite{High12}. The polarization texture in the external ring region appears as the superposition of the polarization textures produced by the MOES beads with each being similar to the texture produced by an LBS (Fig. 1d,e). A periodic array of beads in the MOES (Fig. 1c) creates periodic polarization textures (Fig. 1d,e). The
periodic polarization textures in the external ring region vanish above the critical temperature of the MOES \cite{SI}.

Below we present a theoretical model which describes the appearance of the exciton polarization textures and links them to
spin currents carried by electrons and holes bound into bright and dark exciton states. This model is based on ballistic exciton transport out of the LBS origin and coherent precession of spins of electrons and holes. The former originates from the suppression of scattering and the latter from the suppression of spin relaxation in the condensate of indirect excitons. The states with different spins are split due to the splitting of linearly polarized exciton states and SO interaction, which is described by the Dresselhaus Hamiltonian
$H_{e}=\beta_{e}\left(k_{x}^{e}\sigma_{x}-k_{y}^{e}\sigma_{y}\right)$ for electrons and
$H_{h}=\beta_{h}\left(k_{x}^{h}\sigma_{x}+k_{y}^{h}\sigma_{y}\right)$ for holes \cite{Rashba88, Wu08, Luo10} ($\mathbf{k}_{e,h}$ are electron and hole wave-vectors given by $k_e = k_{ex} m_e /(m_e + m_h)$, $k_h = k_{ex} m_h /(m_e + m_h)$, $m_e$ and $m_h$ are in-plane effective masses of electron and heavy hole, respectively, $k_{ex}$ is the exciton wave vector, $\beta_{e,h}$ are constants, and $\sigma _{x,y}$ are Pauli matrices). In the basis of four exciton states with spins $J_z = +1, -1, +2, -2$, the coherent spin dynamics in the system is governed by a model matrix Hamiltonian:
\begin{equation}%
\hat{H} = \left[
 \begin{array}{ c c c c}
    E_b-(g_h-g_e)\mu_BB/2 & -\delta_b & k_e\beta_ee^{-i\phi} & k_h\beta_he^{-i\phi}\\
    -\delta_b & E_b+(g_h-g_e)\mu_BB/2 & k_h\beta_he^{i\phi} &k_e\beta_ee^{i\phi}\\
    k_e\beta_ee^{i\phi} & k_h\beta_he^{-i\phi} & E_d-(g_h+g_e)\mu_BB/2 &-\delta_d\\
    k_h\beta_he^{i\phi} & k_e\beta_ee^{-i\phi} & -\delta_d &E_d+(g_h+g_e)\mu_BB/2
 \end{array} \right].
\end{equation}%
where $E_{b}$ and $E_{d}$ are energies of bright and dark excitons in an ideal isotropic QW, $\delta_{b}$ and $\delta_{d}$ describe the effect of in-plane anisotropy resulting in the splitting of exciton states linearly polarized along the axes of symmetry. The angle $\varphi$ is measured from the $x$ axis. The details of this model are presented in \cite{SI}. Exciton propagation out of the origin governed by this Hamiltonian results in the appearance of a vortex of linear polarization with the polarization perpendicular to the radial direction and a four-leaf pattern of circular polarization in $B=0$, as well as spiral patterns of linear polarization and bell-like patterns of circular polarization in finite magnetic fields. This model qualitatively reproduces the main features of the experiment for both linear (Fig. 2a-d) and circular (Fig. 2e-h) polarizations.

\begin{figure*}[tbp]
\centering
\includegraphics[width=13cm]{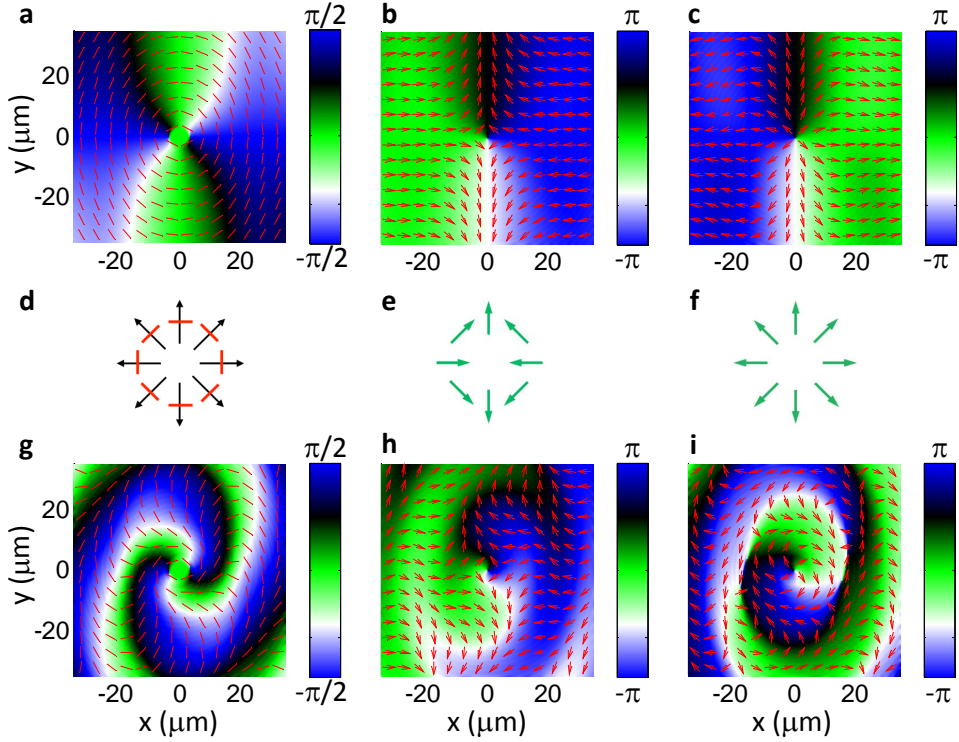}
\caption{{\bf Spin textures.} Simulated in-plane exciton polarization (a), electron spin (b), and hole spin (c) patterns. (d) Schematic of exciton momentum (black arrows) and linear polarization (red lines) patterns. Schematic of effective magnetic fields given by the Dresselhaus SO interaction for electrons (e) and holes (f). Exciton polarization (g), electron spin (h), and hole spin (i) patterns in applied magnetic field $B = 7$ T. The lines (arrows) and the color visualize the orientation of the linear polarization (spin).}
\end{figure*}

This model describes the exciton polarization currents and allows deducing the spin currents carried by electrons and holes bound to excitons as detailed in \cite{SI}. Figure 3b,c show the electron and hole spin textures deduced from the measured exciton polarization texture (Fig. 2a, 3a). One can see that both the electron and hole spin tend to align along the effective magnetic fields given by the Dresselhaus SO interaction ${\bf B}_{eff(e)} = \frac{2 \beta_e}{g_e \mu_B} (- k_{e,x}, k_{e,y})$, ${\bf B}_{eff(h)} = \frac{2 \beta_h}{g_h \mu_B} (k_{h,x}, k_{h,y})$ (Fig. 3e,f), consistent with the model. The patterns of $P_{lin}$ corresponding to the simulations in Fig. 3 are shown in Fig. 2b.

The model can be improved by including non-linear effects. In \cite{SI}, we present simulations of exciton spin currents using Gross-Pitaevskii type equations, which treat the excitons as a coherent field outside the LBS center and include dispersion and interaction. The simulation results are similar to that within the density matrix approach and are in agreement with the experiment. Non-linear spin-related phenomena form interesting perspectives for future studies. In conclusion, long-range spin currents governed by spin-orbit interaction and controlled by applied magnetic field have been observed in a coherent exciton gas.

We thank Misha Fogler, Jorge Hirsch, Leonid Levitov, Yuriy Rubo, Lu Sham, Ben Simons, and Congjun Wu for discussions. This work was supported by DOE. The development of spectroscopy in the dilution refrigerator was also supported by ARO and NSF. AK acknowledges financial support from the Russian Ministry of Education and Science (contract No.  11.G34.31.0067). T.O. was supported by the Ministry of Education and the Grant Agency of the Czech Republic. A.A.H. was supported by an Intel fellowship. J.R.L. was supported by a Chateaubriand Fellowship. T.L. was supported by the EU FP7 Marie Curie EPOQUES project. The collaboration was supported by EU ITN INDEX.

\FloatBarrier

\setcounter{figure}{0}
\setcounter{equation}{0}
\makeatletter
\renewcommand{\thefigure}{S\@arabic\c@figure}

\begin{center}
{\Large \bf Supplementary Materials}
\end{center}

\section{Theory of exciton spin currents: Density matrix approach}

\subsection{Ballistic exciton transport with coherent spin precession}

In zinc-blend semiconductor quantum wells (e.g. in GaAs/AlGaAs system), the lowest energy exciton states are formed by electrons with spin projections to the structure axis of $+1/2$ and $-1/2$ and heavy holes whose quasi-spin (sum of spin and orbital momentum) projection to the structure axis is $+3/2$ or $-3/2$. Consequently, the exciton spin defined as a sum of electron spin and heavy hole quasi-spin may have one of four projections to the structure axis: $-2, -1, +1, +2$ \cite{Ivchenko1997}. These states are usually nearly degenerate, while there may be some splitting between them due to the short and long-range exchange interactions. Here we derive the exciton Hamiltonian in the basis of $+1, -1, +2, -2$ states, accounting for the spin-orbit interaction \cite{Rashba1988, Luo2010}, long- and short-range exchange interactions \cite{Ivchenko2005} and Zeeman effect, but neglecting exciton-exciton interactions. We consider excitons propagating ballistically in plane of a quantum well. We shall characterize them by a fixed wave-vector $k_{ex}$.
	
In order to build the $4 \times 4$ matrix Hamiltonian for excitons, we start with simpler $2 \times 2$ Hamiltonians describing the spin-orbit Dresselhaus effect and Zeeman effect for electrons and holes.

The electron Hamiltonian in the basis of $(+1/2, -1/2)$ spin states writes:
\begin{equation}
H_e= \beta_e (k_{e,x} \sigma_x - k_{e,y} \sigma_y) - \frac{1}{2} g_e \mu_B B \sigma_z .
\end{equation}
Here $g_e$ is the electron $g$-factor, $\mu_B$ is the Bohr magneton, $B$ is a magnetic field normal to the quantum well plane, $\beta_e$ is the Dresselhaus constant describing spin-orbit interactions of electrons, the Pauli matrices are

$ \sigma_x = \left[
 \begin{array}{ c c }
    0 & 1 \\
    1 & 0
 \end{array} \right]
$,
$ \sigma_y = \left[
 \begin{array}{ c c }
    0 & -i \\
    i & 0
 \end{array} \right]
$,
$ \sigma_z = \left[
 \begin{array}{ c c }
    1 & 0 \\
    0 & -1
 \end{array} \right]$.
Hence
\begin{equation}
H_e = \left[
 \begin{array}{ c c }
    -\frac{1}{2}g_e \mu_B B & \beta_e (k_{e,x} + i k_{e,y}) \\
    \beta_e (k_{e,x} - i k_{e,y}) & \frac{1}{2}g_e \mu_B B
 \end{array} \right]
 = \left[
 \begin{array}{ c c }
    -\frac{1}{2}g_e \mu_B B & \beta_e k_ee^{i\phi} \\
    \beta_e k_ee^{-i\phi} & \frac{1}{2}g_e \mu_B B
 \end{array} \right].
\end{equation}
Here $\phi$ is the angle between the electron wave-vector ${\bf k_e}$ and the chosen $x$-axis. The exciton Hamiltonian needs to be written in the basis of $(+1, -1, +2, -2)$ exciton states, which correspond to $(-1/2, +1/2, +1/2, -1/2)$ electron states. The electron spin-flip couples $+1$ and $+2$ states and $-1$ and $-2$ states. For each of these two couples of states we apply the Hamiltonian (2), which results in the following electronic contribution to the $4 \times 4$ exciton Hamiltonian:
\begin{equation}
\widehat{H_e} = \left[
 \begin{array}{ c c c c}
    g_e \mu_B B/2 & 0 & k_e \beta_e e^{-i \phi} & 0\\
    0 & -g_e \mu_B B/2 & 0 &k_e \beta_e e^{i \phi}\\
    k_e \beta_e e^{i \phi} & 0 & -g_e \mu_B B/2 &0\\
    0 & k_e \beta_e e^{-i \phi} & 0 &g_e \mu_B B/2\\
 \end{array} \right].
\end{equation}

The similar reasoning applies to the heavy hole contribution to the Hamiltonian: The hole Hamiltonian written in the basis of $(+3/2,-3/2)$ states is
\begin{equation*}
H_h = \beta_h(k_{h,x} \sigma_x + k_{h,y}\sigma_y) - \frac{1}{2}g_h \mu_B B \sigma_z.
\end{equation*}
Here $g_h$ is the heavy hole $g$-factor, $\beta_h$ is the Dresselhaus constant for heavy holes \cite{Rashba1988, Luo2010}.
Hence,
\begin{equation}
H_h = \left[
 \begin{array}{ c c }
    -\frac{1}{2}g_h \mu_B B & \beta_h (k_{h,x} - i k_{h,y}) \\
    \beta_h (k_{h,x} + i k_{h,y}) & \frac{1}{2}g_h \mu_B B
 \end{array} \right]
= \left[
 \begin{array}{ c c }
    -\frac{1}{2}g_h \mu_B B & \beta_h k_h e^{-i \phi} \\
    \beta_h k_h e^{i \phi} & \frac{1}{2}g_h \mu_B B
 \end{array} \right].
\end{equation}
Here $\phi$ is the angle between the hole wave vector $k_h$ and the chosen $x$-axis. The exciton Hamiltonian is written in the basis of $(+1, -1, +2, -2)$ exciton states, which correspond to $(+3/2, -3/2, +3/2, -3/2)$ hole states. The hole spin-flip couples $+1$ and $-2$ states and $-1$ and $+2$ states. For each of these two couples of states we apply the Hamiltonian (4), which results in the following hole contribution to the $4 \times 4$ exciton Hamiltonian:

\begin{equation}
\widehat{H_h} = \left[
 \begin{array}{ c c c c}
    -g_h \mu_B B/2 & 0 & 0 & k_h \beta_h e^{-i \phi} \\
    0 & g_h \mu_B B/2 & k_h \beta_h e^{i \phi} &0\\
    0 & k_h \beta_h e^{-i \phi} & -g_h \mu_B B/2 &0\\
    k_h \beta_h e^{i \phi} & 0 & 0 &g_h \mu_B B/2\
 \end{array} \right].
\end{equation}

For the translational motion of an exciton as a whole particle the exciton momentum is given by ${\bf P}_{ex}=(m_e+m_{hh} ) {\bf v}_{ex}$, where $m_e$ and $m_{hh}$ are in-plane effective masses of an electron and of a heavy hole, respectively, $v_{ex}$  is the exciton speed. Having in mind that the exciton translational momentum is a sum of electron and hole translational momenta given by ${\bf P}_{e,h} = m_{e,hh} {\bf v}_{e,h}$, ${\bf v}_{e,h}$ being the electron (hole) speed, one can easily see that ${\bf v}_h={\bf v}_e = {\bf v}_{ex}$. Having in mind that ${\bf P}_{ex} = \hbar {\bf k}_{ex}$, ${\bf P}_{e,h} = \hbar {\bf k}_{e,h}$ we have ${\bf k}_{ex} = {\bf k}_h+{\bf k}_e$, $k_e = \frac{m_e}{m_e+m_{hh}} k_{ex}$, $k_h = \frac{m_{hh}}{m_e+m_{hh}} k_{ex}$.\\

Besides the contributions from electron and hole spin orbit interactions and Zeeman splitting, there may be a purely excitonic contribution to the Hamiltonian, which is composed from the Hamiltonian for bright excitons written in the basis $(+1, -1)$:
\begin{equation}
H_b = E_bI- \delta_b \sigma_x = \left[
 \begin{array}{ c c }
    E_b & -\delta_b \\
    -\delta_b & E_b
 \end{array} \right],
\end{equation}
and the Hamiltonian for dark excitons written in the basis $(+2, -2)$:
\begin{equation}
H_d = E_dI- \delta_d \sigma_x = \left[
 \begin{array}{ c c }
    E_d & -\delta_d \\
    -\delta_d & E_d
 \end{array} \right],
\end{equation}
where $I$ is the identity matrix. The terms with $\delta_b$ and $\delta_d$ describe the splittings of bright and dark states polarized along $x$ and $y$ axes in the plane of the structure due to the long-range exchange interaction. Such splitting may appear due to some in-plane anisotropy in the structure induced by strain or monolayer fluctuations of interfaces. We assume that it induced the splitting of $X$- and $Y$-polarized excitons, while it can be easily generalized to the splitting in diagonal or random axes. $E_b - E_d$ is the splitting between bright $+1$ and $-1$ and dark $+2$ and $-2$ exciton states which may be also split due to the short range exchange interaction. Note that Eqs. (6,7) can be simply obtained from the exciton Hamiltonians written in the basis of $(X,Y)$ polarizations. E.g. for the bright excitons:
\begin{equation}
H_{XY} = \left[
 \begin{array}{ c c }
    E_b - \delta_b & 0 \\
    0 & E_b + \delta_b
 \end{array} \right],
\end{equation}
\begin{equation}
\nonumber
H_b=C^{-1}H_{XY}C,
\end{equation}
where
$C= \frac{1}{\sqrt{2}}\left[
 \begin{array}{ c c }
    1 & 1 \\
    i & -i
 \end{array} \right],$
 $C^{-1}= \frac{1}{\sqrt{2}}\left[
 \begin{array}{ c c }
    1 & -i \\
    1 & i
 \end{array} \right]$
are the transformation matrices from linear to circular basis and vice versa \cite{Born1970}. The same reasoning applies to the dark excitons as well. The sum of Hamiltonians (6,7) in the $4 \times 4$ basis writes:

\begin{equation}
H_0 = \left[
 \begin{array}{ c c c c}
    E_b & -\delta_b & 0 & 0\\
    -\delta_b & E_b & 0 &0\\
    0 & 0 & E_d &-\delta_d\\
    0 & 0 & -\delta_d &E_d
 \end{array} \right].
\end{equation}

Let us consider the excitons propagating with a wavevector ${\bf k}_{ex}$. We shall describe them by a spin density matrix $\hat{p}=|\Psi >< \Psi |$, where $\Psi=(\Psi_{+1},\Psi_{-1},\Psi_{+2},\Psi_{-2})$ is the exciton wave-function projected to four spin states. The elements of this density matrix $\rho_{ij}$ are dependent on the distance from the excitation spot $r = v_{ex} t$ and the polar angle $\phi$. The elements of the upper left quarter of the density matrix are linked to the intensity of light emitted by bright exciton states I and to the components of the Stokes vector $S_x$, $S_y$ and $S_z$ of the emitted light:
\begin{equation}
\rho_{11}= \frac{1}{2}+S_z, \rho_{12}=S_x-iS_y,\rho_{21}=S_x+iS_y,\rho_{22}=\frac{1}{2}-S_z.
\end{equation}

These expressions can be summarized using the Pauli matrices as $\left[
\begin{array}{ c c }
    \rho_{11} & \rho_{12} \\
    \rho_{21} & \rho_{22}
 \end{array} \right]=\frac{1}{2} \hat{I} + {\bf S}\hat{{\bf \sigma}}$, where $\hat{I}$ is the identity matrix.

The components of the Stokes vector are directly proportional to the polarization degree of light measured in $XY$ axes, diagonal axes and the circular basis. The circular polarization degree of light emitted by propagating excitons can be obtained as
\begin{equation}
\rho_c = 2S_z=(\rho_{11}-\rho_{22})/(\rho_{11}+\rho_{22}),
\end{equation}
the linear polarization degree can be found from
\begin{equation}
\rho_l = 2S_x=(\rho_{12}+\rho_{21})/(\rho_{11}+\rho_{22}),
\end{equation}
the linear polarization degree measured in the diagonal axes (also referred to as a diagonal polarization degree) is given by
\begin{equation}
\rho_d = 2S_y=i(\rho_{12}-\rho_{21})/(\rho_{11}+\rho_{22}).
\end{equation}
The dynamics of this density matrix is given by the quantum Liouville equation:
\begin{equation}
i\hbar \frac{d\hat{\rho}}{dt}=[\widehat{H}, \hat{\rho}],
\end{equation}
where the Hamiltonian is composed from the electron, hole and exciton contributions given by Eqs. (3,5,9) as follows:
\begin{equation}
\hat{H} = \left[
 \begin{array}{ c c c c}
    E_b-(g_h-g_e)\mu_BB/2 & -\delta_b & k_e\beta_ee^{-i\phi} & k_h\beta_he^{-i\phi}\\
    -\delta_b & E_b+(g_h-g_e)\mu_BB/2 & k_h\beta_he^{i\phi} &k_e\beta_ee^{i\phi}\\
    k_e\beta_ee^{i\phi} & k_h\beta_he^{-i\phi} & E_d-(g_h+g_e)\mu_BB/2 &-\delta_d\\
    k_h\beta_he^{i\phi} & k_e\beta_ee^{-i\phi} & -\delta_d &E_d+(g_h+g_e)\mu_BB/2
 \end{array} \right].
\end{equation}

The Hamiltonian (15) includes the electron, hole, and exciton contributions. Magnetic field affects the electron and hole contributions via the Zeeman splitting. Its effect on the exciton contributions, originating from the change of short and long-range exchange interactions, is neglected in the model. A qualitative agreement with the experiment justifies this approximation.

Note, that in a similar way one can describe the Rashba effect for electrons and holes on the exciton spin density matrix. Our estimations show that for the value of bias we use in these experiments the Rashba effect is much weaker than the Dresselhaus effect \cite{Leonard2009}. Therefore we limit ourself to the consideration of the Dresselhaus effect for electrons and holes. In order to make sure that the observed exciton polarization textures are indeed governed by the Dresselhaus effect, we have performed also the simulations accounting for the Rashba instead of Dresselhaus mechanism of spin-orbit coupling. These simulations produce exciton polarization patterns qualitatively different from the experimental data.

\FloatBarrier
\subsection{An example of how the Dresselhaus effect affects the polarization of propagating exciton}

In order to obtain the spatial distribution of Stokes vector components in the cw regime we assume that all excitons propagate in radial directions from a point-like or a ring-like source. Their polarization state in a point characterized by the polar coordinates $(r,\phi)$ is readily obtained from the elements of the density matrix $\hat{\rho}(t,\phi)$ with $t = r/v_{ex}$. The exciton speed $v_{ex}$ governs the spatial scale of the polarization textures.

Let us consider the simplest example of how the Dresselhaus effect affects the polarization of propagating excitons. In order to do it, we shall commute both parts in Eq. (14) with the Hamiltonian. As a result we shall have:
\begin{equation}
i\hbar \frac{d[\widehat{H},\hat{\rho}]}{dt}=[\hat{H},[\widehat{H},\hat{\rho}]].
\end{equation}

\noindent Now we take a time derivative from both parts of Eq. (14) and substitute the expression (16) in its right part:
\begin{equation}
-\hbar^2 \frac{d^2\hat{\rho}}{dt^2}=[\hat{H},[\widehat{H},\hat{\rho}]].
\end{equation}
Let us suppose that initially we have an exciton state composed by bright excitons linearly polarized along $x$-axis and dark excitons linearly polarized along $y$-axis. It is described by the density matrix:
\begin{equation}
\nonumber
\hat{\rho}_0 = \left[
 \begin{array}{ c c c c}
    1 & 1 & 0 & 0\\
    1 & 1 & 0 &0\\
    0 & 0 & 1 &-1\\
    0 & 0 & -1 &1
 \end{array} \right].
\end{equation}
How this matrix would evolve in time due to the Dresselhaus effect on the electron spin? Let us assume $B=0$ for simplicity, and calculate the commutator of the Hamiltonian (3)
\begin{equation}
\nonumber
\widehat{H_e} = \left[
 \begin{array}{ c c c c}
    0 & 0 & k_e\beta_ee^{-i\phi} & 0\\
    0 & 0 & 0 &k_e\beta_ee^{i\phi}\\
    k_e\beta_ee^{i\phi} & 0 & 0 &0\\
    0 & k_e\beta_ee^{-i\phi} & 0 &0
 \end{array} \right].
\end{equation}
with the density matrix $\hat{\rho}_0$. One can easily see that
\begin{equation}
\nonumber
[\widehat{H_e},\hat{\rho}_0] = 2k_e\beta_e\left[
 \begin{array}{ c c c c}
    0 & 0 & 0 & -\cos{\phi}\\
    0 & 0 & -\cos{\phi} &0\\
    0 & \cos{\phi} & 0 &0\\
    \cos{\phi}& 0 & 0 &0
 \end{array} \right].
\end{equation}
The double commutator in the right part of Eq. (14) can be now found as:
\begin{equation}
\nonumber
[\widehat{H_e},[\widehat{H_e},\hat{\rho_0}]] = 4(k_e\beta_e)^2\left[
 \begin{array}{ c c c c}
    0 & \cos{\phi}e^{-i\phi} & 0 & 0\\
    \cos{\phi}e^{i\phi} & 0 & 0 &0\\
    0 & 0 & 0 &-\cos{\phi}e^{i\phi}\\
    0& 0 & -\cos{\phi}e^{-i\phi} &0
 \end{array} \right].
\end{equation}
Let us substitute this expression back to Eq. (17) and look at the dynamics of the element $\rho_{12}=S_x-iS_y$ describing the linear and diagonal polarization degrees of bright excitons. One can easily see that $\hbar^2\frac{d^2\rho_{12}}{dt^2}=-\hbar^2\frac{d^2(S_x-iS_y)}{dt^2}=4(k_e\beta_e)^2\cos{\phi}e^{-i\phi}.$
Separating the real and imaginary parts of this equation, we find the dynamics of the Stokes vector components: $\hbar^2\frac{d^2(S_x)}{dt^2}=-4(k_e\beta_e)^2\cos{^2\phi};\hbar^2\frac{d^2(S_y)}{dt^2}=-2(k_e\beta_e)^2\sin{2\phi}$.\\

Our initial conditions are: at $t=0, S_x=S_{x0}>0 , S_y=0.$ One can see that for $\phi=-\pi/2,\pi/2$ the polarization does not change: $S_x=S_{x0}, S_y=0$. For $\phi=0,\pi$, we have $S_y=0$, while $S_x$ decreases and, eventually, inverts its sign. The negative $S_x$ corresponds to $Y$-polarization. For any other value of $\phi$ both $S_x$ and $S_y$ change with time. Namely, $S_x$ decreases, $S_y$ builds up. The sign of $S_y$ is negative (corresponds to the polarization along $(1,-1)$ axis) if $0<\phi<\pi/2$ and $\pi<\phi<3\pi/2$. The sign of $S_y$ is positive (corresponds to the polarization along $(1,1)$ axis) if $\pi/2<\phi<\pi$ and $3\pi/2<\phi<2\pi$. This describes a polarization vortex.

In the same way one can show that the Hamiltonian (9) describing the exchange induced linear polarization splitting converts any linear polarization different from $X$ and $Y$ polarizations to the circular polarization.

\FloatBarrier
\subsection{Spin currents carried by electrons and holes bound into excitons}

It is important to note that the present formalism addresses the spin part of the exciton wavefunction, which is a product of electron and hole spin functions. E.g. the probability to find the exciton in the spin state $+1$ is given by a product of probabilities to find an electron in the spin state $-1/2$ and the heavy hole in the spin state $+3/2$. The four component exciton wave-function
\begin{equation}
\Psi = \left(\Psi_{+1}, \Psi_{-1}, \Psi_{+2}, \Psi_{-2},\right) = \left(\Psi_{e,-1/2}\Psi_{h,+3/2}, \Psi_{e,+1/2}\Psi_{h,-3/2}, \Psi_{e,+1/2}\Psi_{h,+3/2}, \Psi_{e,-1/2}\Psi_{h,-3/2}\right),
\end{equation}
where  $\Psi_{e,+1/2}$ and $\Psi_{e,-1/2}$ are the components of the electron spinor wave function, $\Psi_{h,+3/2}$ and $\Psi_{h,-3/2}$ are the components of the heavy hole spinor wave function. We shall normalize exciton, electron and hole wave functions to 1, namely:
\begin{equation}
\Psi_{+1}\Psi^*_{+1}+\Psi_{-1}\Psi^*_{-1}+\Psi_{+2}\Psi^*_{+2}+\Psi_{-2}\Psi^*_{-2} =
\Psi_{e,+\frac{1}{2}}\Psi^*_{e,+\frac{1}{2}}+\Psi_{e,-\frac{1}{2}}\Psi^*_{e,-\frac{1}{2}} =
\Psi_{h,+\frac{3}{2}}\Psi^*_{h,+\frac{3}{2}}+\Psi_{h,-\frac{3}{2}}\Psi^*_{h,-\frac{3}{2}} = 1.
\end{equation}
Now, the exciton spin density matrix is given by
\begin{equation*}
\hat{\rho} = \left|\Psi\right>\left<\Psi\right|
=\left[\begin{array}{cccc}
\Psi_{+1}\Psi^*_{+1} & \Psi_{+1}\Psi^*_{-1} & \Psi_{+1}\Psi^*_{+2} & \Psi_{+1}\Psi^*_{-2}\\
\Psi_{-1}\Psi^*_{+1} & \Psi_{-1}\Psi^*_{-1} & \Psi_{-1}\Psi^*_{+2} & \Psi_{-1}\Psi^*_{-2}\\
\Psi_{+2}\Psi^*_{+1} & \Psi_{+2}\Psi^*_{-1} & \Psi_{+2}\Psi^*_{+2} & \Psi_{+2}\Psi^*_{-2}\\
\Psi_{-2}\Psi^*_{+1} & \Psi_{-2}\Psi^*_{-1} & \Psi_{-2}\Psi^*_{+2} & \Psi_{-2}\Psi^*_{-2}\end{array}\right] =
\end{equation*}
\begin{equation}
=\left[\begin{array}{cccc}
\Psi^{}_{e,-\frac{1}{2}}\Psi^*_{e,-\frac{1}{2}}\Psi^{}_{h,+\frac{3}{2}}\Psi^*_{h,+\frac{3}{2}} &
\Psi^{}_{e,-\frac{1}{2}}\Psi^*_{e,+\frac{1}{2}}\Psi^{}_{h,+\frac{3}{2}}\Psi^*_{h,-\frac{3}{2}} &
\Psi^{}_{e,-\frac{1}{2}}\Psi^*_{e,+\frac{1}{2}}\Psi^{}_{h,+\frac{3}{2}}\Psi^*_{h,+\frac{3}{2}} &
\Psi^{}_{e,-\frac{1}{2}}\Psi^*_{e,-\frac{1}{2}}\Psi^{}_{h,+\frac{3}{2}}\Psi^*_{h,-\frac{3}{2}}\\
\Psi^{}_{e,+\frac{1}{2}}\Psi^*_{e,-\frac{1}{2}}\Psi^{}_{h,-\frac{3}{2}}\Psi^*_{h,+\frac{3}{2}} &
\Psi^{}_{e,+\frac{1}{2}}\Psi^*_{e,+\frac{1}{2}}\Psi^{}_{h,-\frac{3}{2}}\Psi^*_{h,-\frac{3}{2}} &
\Psi^{}_{e,+\frac{1}{2}}\Psi^*_{e,+\frac{1}{2}}\Psi^{}_{h,-\frac{3}{2}}\Psi^*_{h,+\frac{3}{2}} &
\Psi^{}_{e,+\frac{1}{2}}\Psi^*_{e,-\frac{1}{2}}\Psi^{}_{h,-\frac{3}{2}}\Psi^*_{h,-\frac{3}{2}}\\
\Psi^{}_{e,+\frac{1}{2}}\Psi^*_{e,-\frac{1}{2}}\Psi^{}_{h,+\frac{3}{2}}\Psi^*_{h,+\frac{3}{2}} &
\Psi^{}_{e,+\frac{1}{2}}\Psi^*_{e,+\frac{1}{2}}\Psi^{}_{h,+\frac{3}{2}}\Psi^*_{h,-\frac{3}{2}} &
\Psi^{}_{e,+\frac{1}{2}}\Psi^*_{e,+\frac{1}{2}}\Psi^{}_{h,+\frac{3}{2}}\Psi^*_{h,+\frac{3}{2}} &
\Psi^{}_{e,+\frac{1}{2}}\Psi^*_{e,-\frac{1}{2}}\Psi^{}_{h,+\frac{3}{2}}\Psi^*_{h,-\frac{3}{2}}\\
\Psi^{}_{e,-\frac{1}{2}}\Psi^*_{e,-\frac{1}{2}}\Psi^{}_{h,-\frac{3}{2}}\Psi^*_{h,+\frac{3}{2}} &
\Psi^{}_{e,-\frac{1}{2}}\Psi^*_{e,+\frac{1}{2}}\Psi^{}_{h,-\frac{3}{2}}\Psi^*_{h,-\frac{3}{2}} &
\Psi^{}_{e,-\frac{1}{2}}\Psi^*_{e,+\frac{1}{2}}\Psi^{}_{h,-\frac{3}{2}}\Psi^*_{h,+\frac{3}{2}} &
\Psi^{}_{e,-\frac{1}{2}}\Psi^*_{e,-\frac{1}{2}}\Psi^{}_{h,-\frac{3}{2}}\Psi^*_{h,-\frac{3}{2}}\end{array}\right].
\end{equation}
This representation allows us to obtain useful links between the elements of exciton, electron and hole density matrices, in particular:
\begin{equation}
\hat{\rho}_e=\left|\Psi_e\right>\left<\Psi_e\right|=\left[
\begin{array}{cc}
\Psi_{e,+\frac{1}{2}}^{} \Psi_{e,+\frac{1}{2}}^* & \Psi_{e,+\frac{1}{2}}^{} \Psi_{e,-\frac{1}{2}}^*\\
\Psi_{e,-\frac{1}{2}}^{} \Psi_{e,+\frac{1}{2}}^* & \Psi_{e,-\frac{1}{2}}^{} \Psi_{e,-\frac{1}{2}}^* \end{array}
\right]=\left[
\begin{array}{cc}
\rho_{22}+\rho_{33} & \rho_{24}+\rho_{31}\\
\rho_{13}+\rho_{42} & \rho_{11}+\rho_{44}
\end{array}
\right],
\end{equation}
\begin{equation}
\hat{\rho}_h=\left|\Psi_h\right>\left<\Psi_h\right|=\left[
\begin{array}{cc}
\Psi_{h,+\frac{3}{2}}^{} \Psi_{h,+\frac{3}{2}}^* & \Psi_{h,+\frac{3}{2}}^{} \Psi_{h,-\frac{3}{2}}^*\\
\Psi_{h,-\frac{3}{2}}^{} \Psi_{h,+\frac{3}{2}}^* & \Psi_{h,-\frac{3}{2}}^{} \Psi_{h,-\frac{3}{2}}^* \end{array}
\right]=\left[
\begin{array}{cc}
\rho_{11}+\rho_{33} & \rho_{14}+\rho_{32}\\
\rho_{23}+\rho_{41} & \rho_{22}+\rho_{44}
\end{array}
\right].
\end{equation}
We know that the components of electron and hole density matrices are linked with the projections of electron and hole spins as:
\begin{equation}
\hat{\rho}_e = \left[
\begin{array}{cc}
\frac{1}{2} + S_{e,z} & S_{e,x} - iS_{e,y}\\
S_{e,x} + iS_{e,y} & \frac{1}{2} - S_{e,z}
\end{array}
\right],
\hat{\rho}_h = \left[
\begin{array}{cc}
\frac{1}{2} + S_{h,z} & S_{h,x} - iS_{h,y}\\
S_{h,x} + iS_{h,y} & \frac{1}{2} - S_{h,z}
\end{array}
\right],
\end{equation}
(here, for the heavy hole we have assigned spin $+1/2$ to the state $+3/2$ and spin $-1/2$ to the state $-3/2$ accounting for the orbital momentum of these states of $+1$ and $-1$, respectively). Therefore, the $z$-component of spin polarization carried by electrons can be expressed as
\begin{equation}
S_{e,z}=(\rho_{22}+\rho_{33}-\rho_{11}-\rho_{44})/2,
\end{equation}
and the $z$-component of spin polarization carried by holes can be expressed as
\begin{equation}
S_{h,z}=(\rho_{11}+\rho_{33}-\rho_{22}-\rho_{44})/2.
\end{equation}

The in-plane component of electron and hole spins can be extracted from the off-diagonal elements of the density matrix. Namely, the $x$-component of electron spin is given by
\begin{equation}
S_{e,x}=(\rho_{13}+\rho_{31}+\rho_{24}+\rho_{42})/2,
\end{equation}
the $x$-component of the hole spin is given by
\begin{equation}
S_{h,x}=(\rho_{14}+\rho_{23}+\rho_{32}+\rho_{41})/2,
\end{equation}
the $y$-component of electron spin is given by
\begin{equation}
S_{e,y}=i(-\rho_{13}+\rho_{31}+\rho_{24}-\rho_{42})/2,
\end{equation}
and the $y$-component of the hole spin is given by
\begin{equation}
S_{h,y}=i(\rho_{14}-\rho_{23}+\rho_{32}-\rho_{41})/2.
\end{equation}

\FloatBarrier
\subsection{Effective fields}

Spin-orbit effect for electrons and holes can be accounted by introducing an in-plane effective magnetic field:
\begin{equation}
H_{eff(e,h)}= -\frac{1}{2} g_{e,h} \mu_B \left({\bf B}_{eff} \hat{\sigma}\right).
\end{equation}
In the case of Dresselhaus Hamiltonian for electrons:
\begin{equation}
-\frac{1}{2} g_{e} \mu_B {\bf B}_{eff} =\beta_e \left(k_{e,x},-k_{e,y} \right).
\end{equation}
For holes the field is collinear to the wave-vector:
\begin{equation}
-\frac{1}{2} g_{h} \mu_B {\bf B}_{eff}=\beta_h \left(k_{h,x},k_{h,y} \right).
\end{equation}

\begin{figure*}[!htbp]
\centering
\includegraphics[width=17cm]{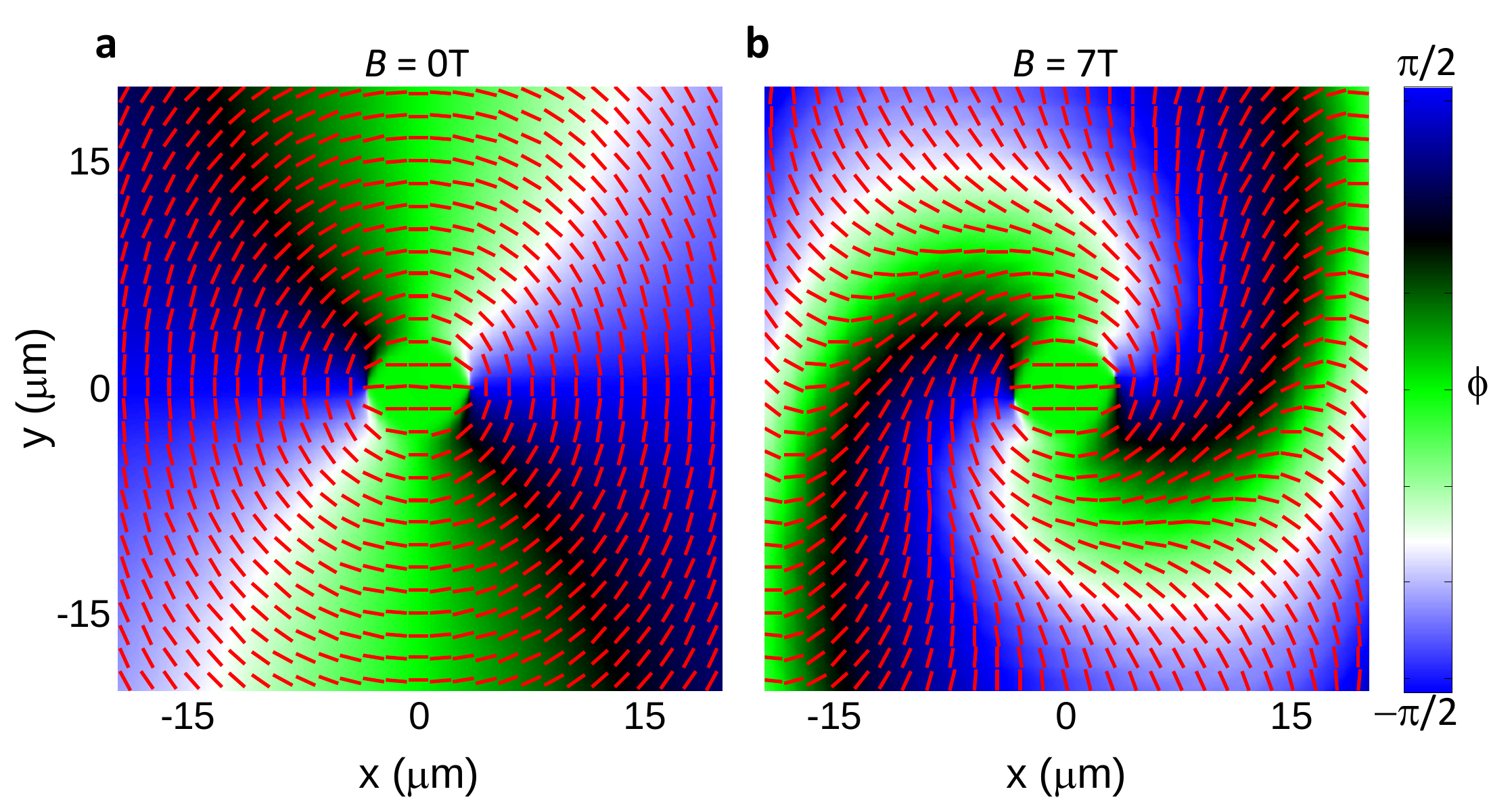}
\caption{{\bf Simulated polarization patterns.} Simulated in-plane exciton polarization in magnetic field $B = 0$ (a) and 7 T (b). The lines and the color visualize the orientation of the linear polarization.}
\end{figure*}

\FloatBarrier
\subsection{Simulations}

The system is modeled using Eq. (15) and parameters are adjusted to match the experimental data (see Fig. 2 in the main text). The same set of parameters is used for linear and circular polarizations and for all magnetic fields. The parameters are $\beta_e = 2.7$ ${\mu}$eV${\mu}$m, $\beta_h = 0.92$ ${\mu}$eV${\mu}$m, $\delta_b = 0.5$ ${\mu}$eV, $\delta_d = -13$ ${\mu}$eV, $E_b - E_d = 5$ ${\mu}$eV, $k_{ex} = 15.4$ ${\mu}$m$^{-1}$, $T = 0.1$ K, $g_e = -0.01$, and $g_h = +0.0085$. Some of the parameters were obtained in earlier studies and some of them were put as  fitting parameters. All fitting parameters were chosen to be consistent with the data published in earlier studies: $\beta_e$ is consistent with our earlier measurements \cite{Leonard2009}, $\beta_h$ -- with the calculated values \cite{Rashba1988, Luo2010}, $T$ -- with the calculated temperature of indirect excitons for the studied structure \cite{Butov2001}, exciton splittings -- with typical exciton splittings in GaAs structures \cite{Leyder2007}, $g$-factors -- with typical $g$-factors in quantum wells with $\sim 8$ nm width \cite{Snelling1992}, and $k_{ex}$ was taken within the light cone in the structure (we checked that the model leads to qualitatively similar patterns for various $k_{ex}$).

The images are presented with spatial averaging over $1.5$ $\mu$m. The initial exciton state considered by this model is a ring around the LBS center where the exciton gas is classical. The ring radius is taken 4 ${\mu}$m. On this ring, the simulations consider the classical exciton energy distribution with $T = 0.1$ K, $k_{ex} = 0$. No simulations were performed inside this ring, the polarization in the ring is shown there. Beyond this ring, the exciton gas is coherent and the simulations consider ballistic exciton transport with coherent spin precession.

Figure S1 (an enlarged version of Fig. 3a,g in the main text) presents simulated patterns of in-plane projection of the Stokes vector of emitted light which directly maps the pseudospin of bright excitons (see e.g. \cite{Kavokin05, Kavokin07}. The patterns of light polarization corresponding to these simulations are shown in Fig. 2b in the main text. Figure S1 visualizes a vortex (a) and a spiral (b) pattern of linear polarization in zero (a) and a finite (b) magnetic field. The simulations produce the observed exciton polarization textures as described in the main text.

\section{Theory of exciton spin currents: Gross-Pitaevskii equations}

\begin{figure*}[!htbp]
\centering
\includegraphics[width=\textwidth]{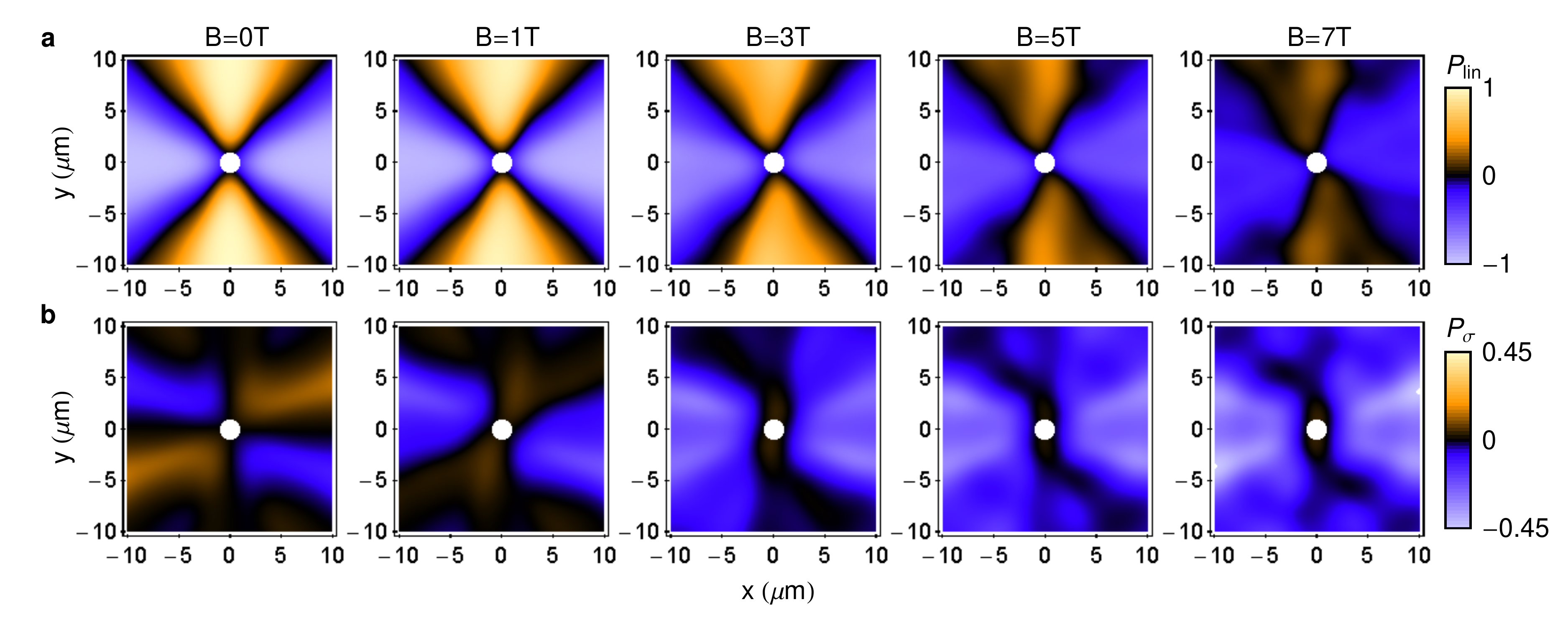}
\caption{{\bf Gross-Pitaevskii simulation.} Spatial distribution of the linear (a) and circular (b) polarization degrees vs magnetic field. As in the case of the density matrix calculations, the results are spatially averaged over $1.5\mu$m. Parameters: $m_{ex}=0.21 m_e$, $\beta_e=10\mu$eV$\mu$m; $\beta_h=\beta_e/3$; $\delta_b=2\mu$eV; $\delta_d=0.5\mu$eV; $E_b-E_d=2\mu$eV; $W=0.2\alpha$, $\alpha n=1\mu$eV. The source area was taken circular with a radius of $1\mu$m.}
\label{fig:GP}
\end{figure*}

The spin density matrix theory is convenient for the description of (partially) coherent and (partially) polarized exciton gases. However, the treatment of non-linear effects in a partially coherent system is a challenging task. An approach, which we consider in this section is to assume a perfectly coherent condensate of excitons. In this case, the excitons can be described by a spatially dependent four-component wavefunction, $\left(\psi_{+1}(\mathbf{x}),\psi_{-1}(\mathbf{x}),\psi_{+2}(\mathbf{x}),\psi_{-2}(\mathbf{x})\right)$. The dynamics of the wavefunction can be described by the Gross-Pitaevskii equation:
\begin{equation}
i\hbar\frac{d}{dt}\left(\begin{array}{c}\psi_{+1}(\mathbf{x})\\\psi_{-1}(\mathbf{x})\\\psi_{+2}(\mathbf{x})\\\psi_{-2}(\mathbf{x})\end{array}\right)=\left(\hat{H}-\frac{\hbar^2\hat{\nabla}^2}{2m_{ex}}\right)
\left(\begin{array}{c}\psi_{+1}(\mathbf{x})\\\psi_{-1}(\mathbf{x})\\\psi_{+2}(\mathbf{x})\\\psi_{-2}(\mathbf{x})\end{array}\right)+\alpha n(\mathbf{x}) \left(\begin{array}{c}\psi_{+1}(\mathbf{x})\\\psi_{-1}(\mathbf{x})\\\psi_{+2}(\mathbf{x})\\\psi_{-2}(\mathbf{x})\end{array}\right)
+W\left(\begin{array}{c}\psi^*_{-1}(\mathbf{x})\psi_{+2}(\mathbf{x})\psi_{-2}(\mathbf{x})\\\psi^*_{+1}(\mathbf{x})\psi_{+2}(\mathbf{x})\psi_{-2}(\mathbf{x})\\\psi^*_{-2}(\mathbf{x})\psi_{+1}(\mathbf{x})\psi_{-1}(\mathbf{x})\\\psi_{+2}(\mathbf{x})\psi_{+1}(\mathbf{x})\psi_{-1}(\mathbf{x})\end{array}\right).\label{eq:GP}
\end{equation}

Here $\hat{H}$ is the same Hamiltonian (15), although to correctly define the operation of the $k-$dependent spin-orbit terms one should replace $k_e=\frac{m_e}{me+m_{hh}}\hat{k}_{ex}$ and $k_h=\frac{m_h}{me+m_{hh}}\hat{k}_{ex}$ where $\hat{k}_{ex}=-i\hat{\nabla}$. Note that the Gross-Pitaevskii equation allows to work with a distribution of wavevectors, and accounts for the dispersion of excitons via the term $-\frac{\hbar^2\hat{\nabla}^2}{2m_{ex}}$, where $m_{ex}$ is the exciton effective mass. The last two terms in Eq.~\ref{eq:GP} are nonlinear terms. In the initial simplified approach presented here we assume a spin-independent scattering rate, $\alpha$, where each spin fraction scatters with the total density, $n(\mathbf{x})=|\psi_{+1}(\mathbf{x})|^2+\psi_{-1}(\mathbf{x})^2+|\psi_{+2}(\mathbf{x})|^2+|\psi_{-2}(\mathbf{x})|^2$. The last term, proportional to $W$, represents a parametric scattering process unique to indirect exciton systems where two bright excitons convert into two dark excitons (or vice versa). Note that all non-linear scattering processes conserve the total spin projection of excitons.

To describe the excitation of our system, we introduce a fixed wavefunction boundary condition along a circular boundary representing the edges of the classical region around the LBS center. We choose a linearly polarized dark exciton density at this boundary, assuming that dark excitons have lower energy than bright excitons. We do not aim to use the Gross-Pitaevskii equation in the hot LBS center where the exiton gas is classical, however, we expect it to offer a qualitative description of the propagation of coherent excitons away from the LBS source in the region beyond the LBS center where the exciton gas is coherent. The indirect excitons have a long lifetime and thus we employ an absorbing boundary condition to allow solution of the problem. The absorbing boundary introduces a loss mechanism in the system, such that a steady state is reached where the excitons excited at the LBS balance the flow of excitons away from the studied region (we assume that upon crossing the absorbing boundary excitons never return).

By solving Eq.~\ref{eq:GP} numerically for the steady state, the linear and circular polarization degrees are calculated for increasing magnetic fields (Fig.~\ref{fig:GP}). The simulated patterns are qualitatively similar to both the experimentally measured patterns and the patterns simulated using the spin density matrix theory.
\FloatBarrier

\section{Experiment}

\subsection{Experimental setup}
\begin{figure*}[!h]
\centering
\includegraphics[width=12cm]{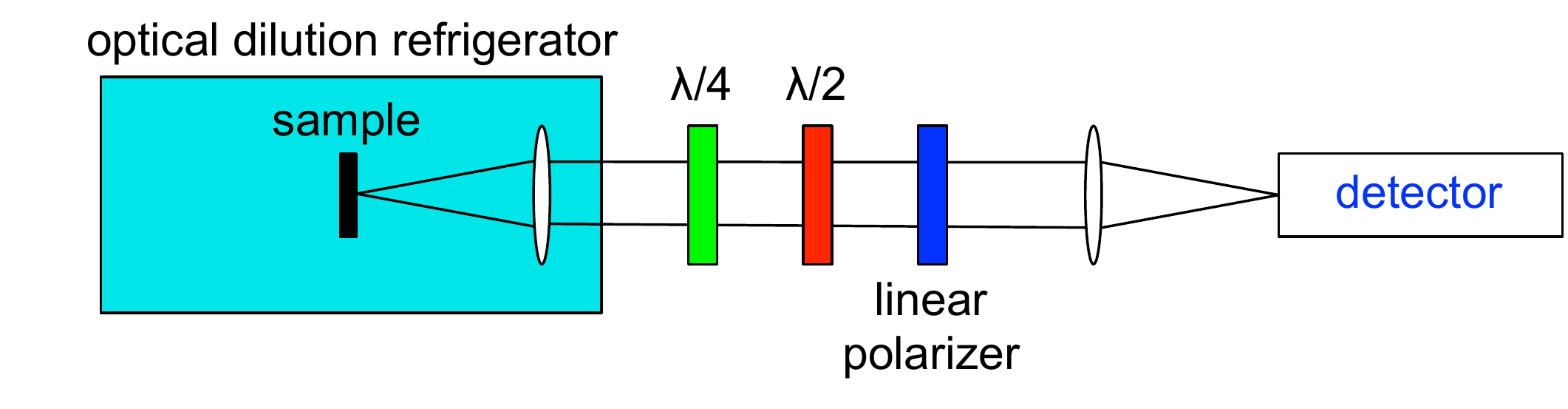}
\caption{{\bf Experimental setup.} Schematic of polarization-resolved imaging.}
\end{figure*}

The schematic of the polarization-resolved imaging experiment is presented in Fig. S3. The sample is in an optical dilution refrigerator. Light emitted by the sample is collected and made parallel by an objective inside the refrigerator. Polarization selection is done by a combination of quarter-wave plate ($\lambda /4$), half-wave plate ($\lambda /2$), and linear polarizer. The linear polarizer is aligned such that $y$-axis polarized emission is transmitted to the detector. The detector is a combination of an interference filter of linewidth $\pm 5$ nm adjusted to the emission wavelength of indirect excitons $\lambda=800$ nm (due to the interference filter, only the emission of indirect excitons is measured: the contribution of the weak emission of direct excitons or any other emission, such as low-energy bulk emission, is cut off by the interference filter), a spectrometer operating in dispersionless zeroth-order mode, and a liquid nitrogen cooled CCD.

\textit{Measurements of linear polarization}. The quarter-wave plate is aligned so that the fast and slow axis are along the $x$- and $y$-axis. Hence, $x$- and $y$-polarized emission $I_x$ and $I_y$ are transmitted unchanged. To measure $I_y$, the half-wave plate fast axis is aligned along the $y$-axis. Then, $I_y$ is transmitted unchanged, and is then transmitted through the linear polarizer. To measure $I_x$, the half-wave plate axis is aligned $45^\text{o}$ relative to the $y$-axis. Then, $I_x$ is rotated to the $y$-axis and is transmitted through the linear polarizer.

\textit{Measurements of circular polarizations}. The quarter-wave plate is aligned so that the fast and slow axis are rotated by $45^\text{o}$ with respect to the $y$-axis. Then, circularly polarized emission $I_{\sigma+}$ and $I_{\sigma-}$ are converted to $x$- and $y$-polarized light. This light is then selected as described above.

In all experiments, the photoexcitation is nonresonant ($> 400$ meV above the energy of indirect excitons) and spatially separated (the 10 $\mu$m-wide excitation spot is $> 80$ $\mu$m away from both the LBS and external ring) so that neither the exciton polarization nor coherence is induced by the pumping light.

\FloatBarrier
\subsection{Patterns of linear polarization and coherence in the LBS ring region}

\begin{figure*}[!htpb]
\includegraphics[width=12cm]{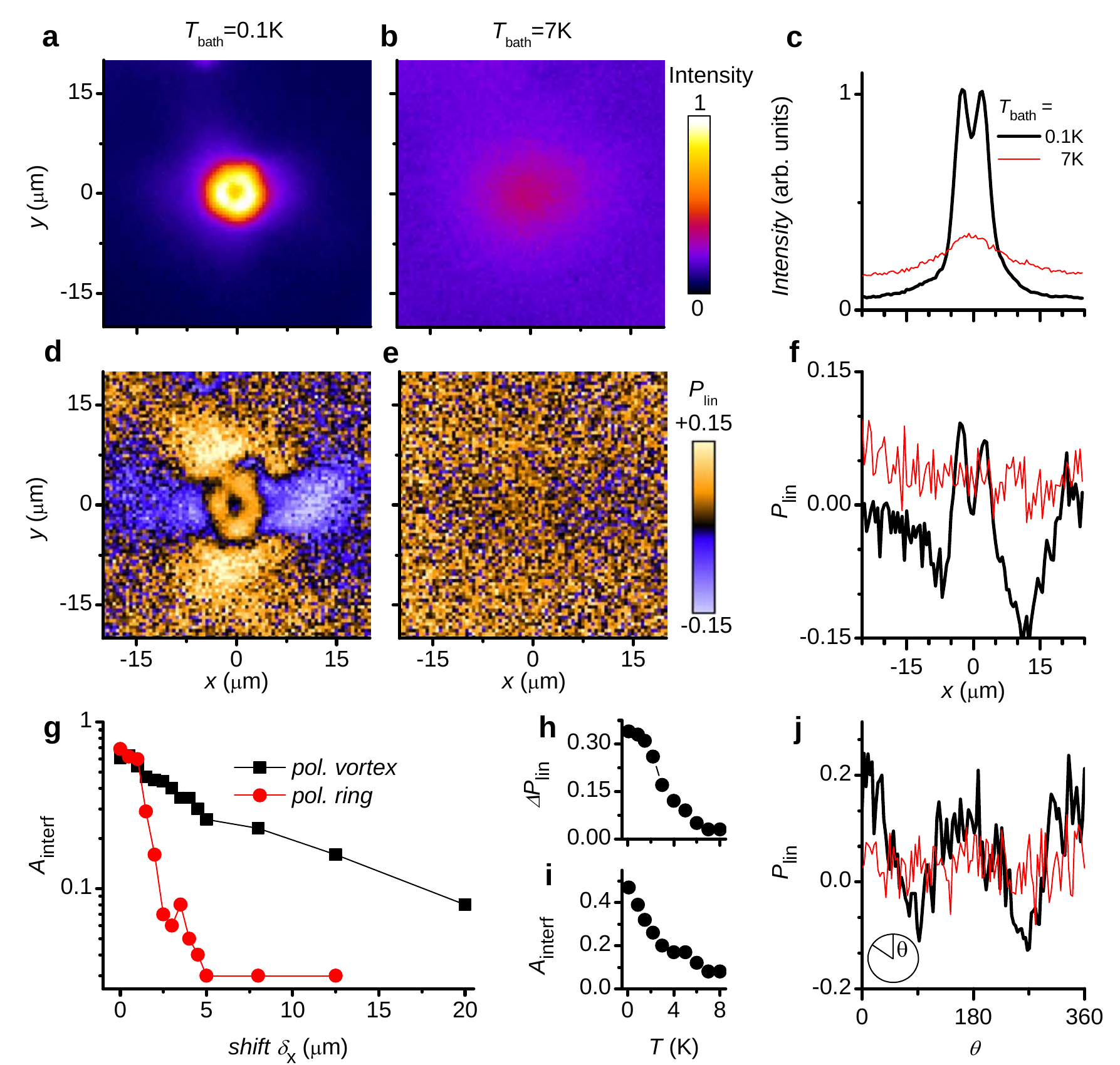}
\caption{{\bf Pattern of linear polarization in the LBS region: temperature dependence.} Measured PL intensity in the region of LBS centered at $(105,75)$ in Fig. 1c at $T_{\protect bath} = 0.1$ (a) and 7 (b) K. (c) PL Intensity profile through the center of the LBS for $T_{\protect bath} = 0.1$ (black) and 7 (red) K. Measured linear polarization of the emission of indirect excitons $P_{\protect lin} = (I_{\protect x} - I_{\protect y}) / (I_{\protect x} + I_{\protect y})$ at $T_{\protect bath} = 0.1$ (d) and 7 (e) K. $P_{\protect lin}$ profile (f) vs $x$ through the LBS center and (j) vs azimuthal angle measured from the $y$-axis at $r=9$ $\protect\mu$m from the LBS center for $T_{\protect bath} = 0.1$ (black) and 7 (red) K. (G) The exciton coherence degree measured by shift-interferometry: Interference visibility $A_{\protect interf}$ vs. shift $\protect\delta_x$ for the vortex of linear polarization (squares), 18 $\protect\mu$m left of LBS center, and the polarization ring (points), 2 $\protect\mu$m left of LBS center for the LBS centered at $(80,105)$ in Fig. 1c. $T_{\protect bath} = 0.1$ K. Emergence of the vortex of linear polarization and spontaneous coherence at low temperatures: (h) The amplitude of azimuthal variation of $P_{\protect lin}$ at $r=10$ $\protect\mu$m from the LBS center and (i) interference visibility $A_{\protect interf}$ in the polarization vortex for $\protect\delta_x = 2$ $\protect\mu$m vs. temperature.}
\end{figure*}

The ring of linear polarization vanishes with increasing temperature (Fig. S4d-f). The vortex of linear polarization vanishes with increasing temperature (Fig. S4d-f,j).

The coherence of an exciton gas is imprinted on the coherence of emission, which is described by the first-order coherence function $g_1(\delta_x)$. In turn, this function is given by the amplitude of the interference fringes $A_{\text{interf}}(\delta_x)$ in `the ideal experiment' with perfect spatial resolution. In real experiments, the measured $A_{\text{interf}}(\delta_x)$ is given by the convolution of $g_1(\delta_x)$ with the point-spread function (PSF) of the optical system used in the experiment \cite{Fogler08}. Both for a classical gas and quantum gas $g_1(\delta_x)$ is close to 1 at $\delta_x = 0$ and drops with increasing $\delta_x$ within the coherence length $\xi$. The difference between the classical and quantum gas is in the value of $\xi$. For a classical gas, $\xi_{cl}$ is close to the thermal de~Broglie wavelength $\lambda_{dB} = \sqrt{\frac{2 \pi \hbar^2}{m T}}$, which is well below the PSF width in the studied temperature range ($\xi_{cl@0.1 K} \sim 0.3~\mu$m, the PSF width is $\sim 1.5~\mu$m). Spontaneous exciton coherence with a large coherence length $\xi$ is observed in the region of the polarization vortex (Fig. S4g, black squares). In contrast, $\xi$ is short in the region of the polarization ring (Fig. S4g, red circles), which is close to the hot LBS center. Large $\xi \gg \xi_{cl}$ in the region of the polarization vortex indicates a coherent exciton state with a much narrower than classical exciton distribution in momentum space, characteristic of a condensate. Small $\xi \sim 1.5$ $\mu$m in the region of the polarization ring indicates a classical exciton state with $\xi$ measuring the PSF width. The coherence measurements are discussed in detail in \cite{High12}. With reducing temperature, the exciton spin textures emerge in concert with coherence (Fig. S4h and S4i).

\FloatBarrier
\subsection{Patterns of circular polarization in the LBS ring region}

\begin{figure*}[!htbp]
\centering
\includegraphics[width=17cm]{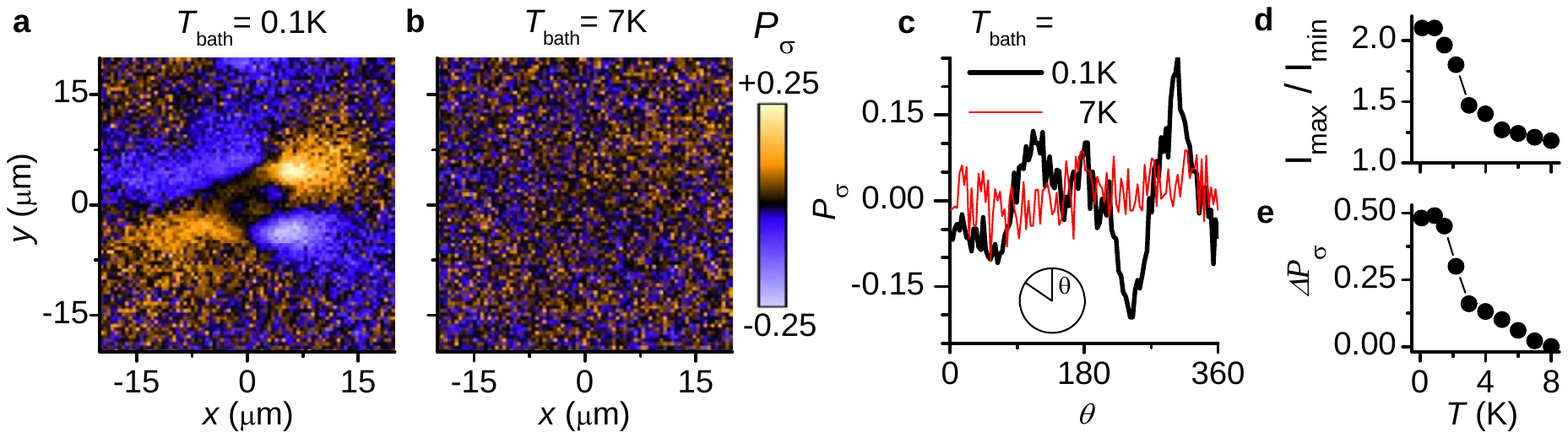}
\caption{{\bf Pattern of circular polarization in the LBS region: temperature dependence.} Measured circular polarization of the emission of indirect excitons $P_{\protect\sigma} = (I_{\protect\sigma^+} - I_{\protect\sigma^-}) / (I_{\protect\sigma^+} + I_{\protect\sigma^-})$ in the LBS region at $T_{\protect bath} = 0.1$ (a) and 7 (b) K. (c) The azimuthal variation of $P_{\protect\sigma}$ at $r=9$ $\mu$m from the LBS center for $T_{\protect bath} = 0.1$ (black) and 7 (red) K. (d) The ratio of maximum to minimum in the azimuthal variation of the total emission intensity of indirect excitons $I_{\protect max}/I_{\protect min}$ at $r = 8 \protect\mu$m from the LBS center vs. temperature. (e) The amplitude of variation of $P_{\protect\sigma}$ around the LBS centered at $(105,75)$ in Fig. 1c, vs. temperature.}
\end{figure*}

The four-leaf pattern of circular polarization vanishes with increasing temperature (Fig. S5a-c,e). At low temperature, the flux of excitons from the LBS center is anisotropic: the emission intensity is enhanced along the polarization direction in the polarization ring (Fig. S6a). To quantify the flux anisotropy, we use the ratio of maximum to minimum in the azimuthal variation of the total emission intensity of indirect excitons $I_{max}/I_{min}$ at distance $r = 8$ $\protect\mu$m from the origin. At high temperature, the exciton flux anisotropy vanishes (Fig. S5d). The four-leaf pattern and flux anisotropy emerge in concert (Fig. S5d and S5e). The four-leaf pattern of circular polarization is associated with a skew of the exciton fluxes in orthogonal circular polarizations (Fig. S6).

\begin{figure*}[!htbp]
\centering
\includegraphics[width=5.5cm]{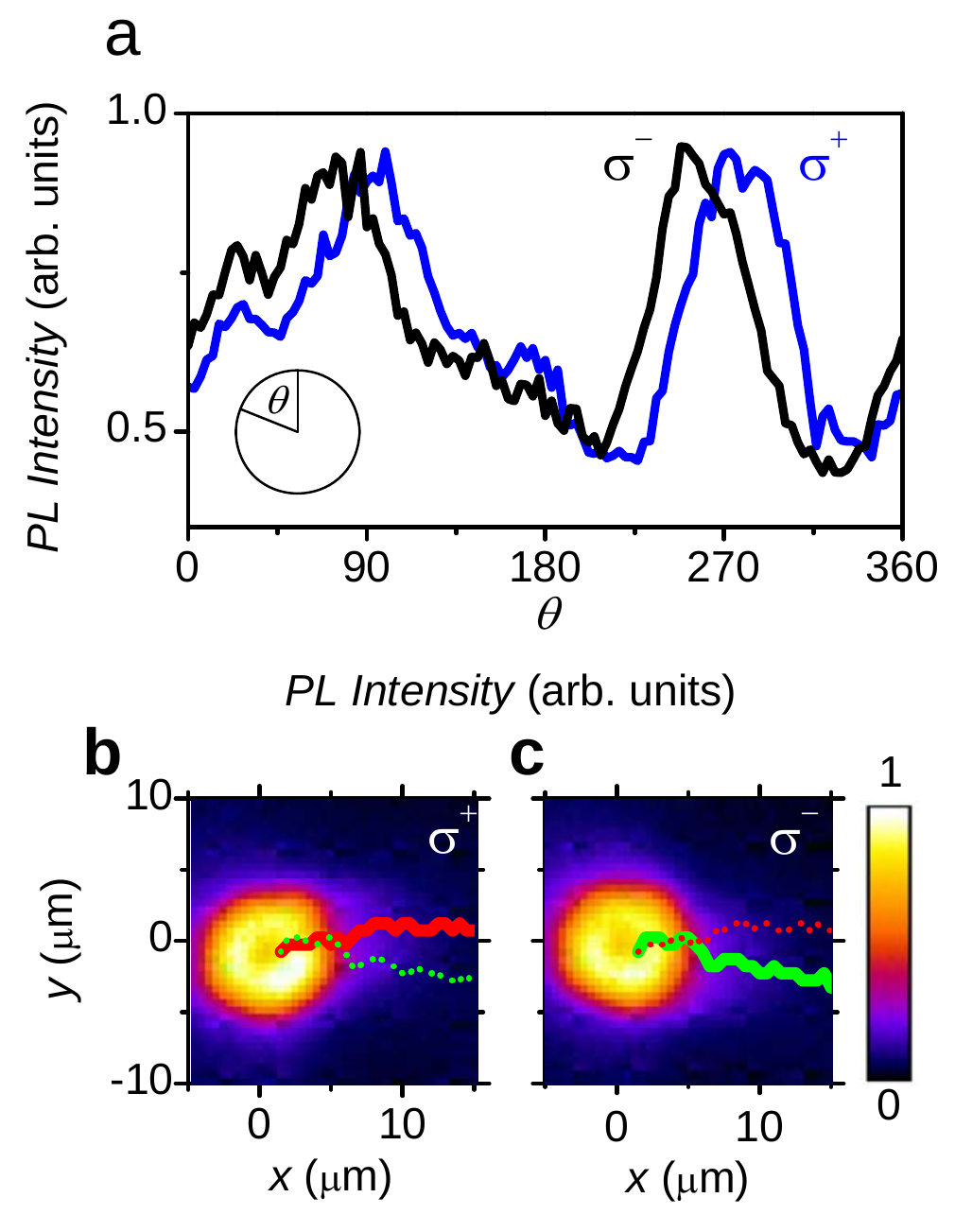}
\caption{{\bf Pattern of circular polarization in the LBS region: azimuthal dependence.} (a) Azimuthal variation of the emission intensity of indirect excitons at $r = 8$ $\protect\mu$m in $\protect\sigma^+$ (blue) and $\protect\sigma^-$ (black) polarizations. Angles are measured from the $y$-axis. (b,c) Traces of the $\protect\sigma^+$ (red) and $\protect\sigma^-$ (green) emission peak around $\protect\theta = 270^{\circ}$ [see (a)]; The emission image in $\protect\sigma^+$ (b) and $\protect\sigma^-$ (c) polarization is also shown.}
\end{figure*}

\FloatBarrier
\subsection{Polarization patterns in the external ring region}

\begin{figure*}[!htbp]
\centering
\includegraphics[width=9.5cm]{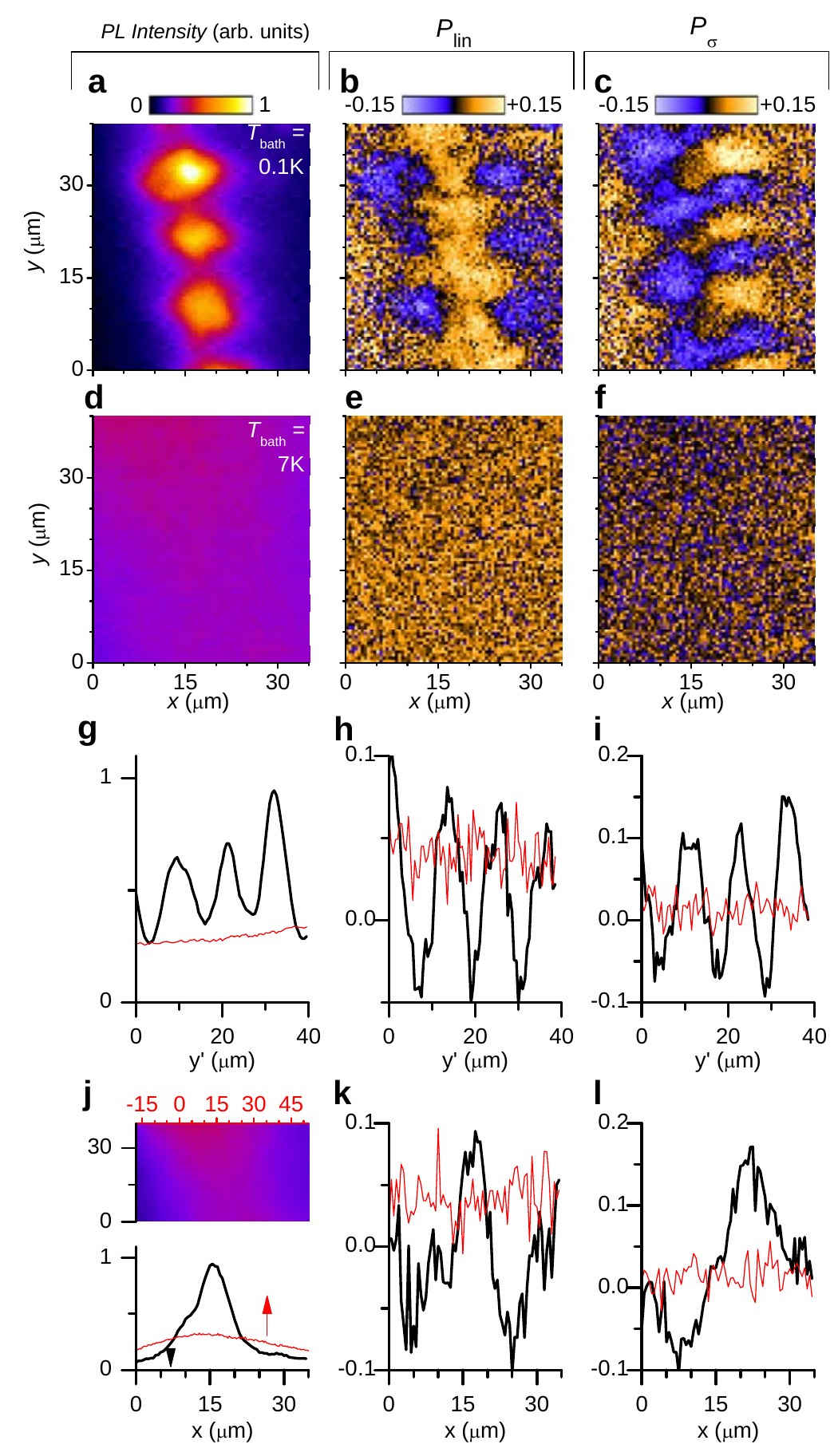}
\caption{{\bf Polarization textures in the region of external ring: temperature dependence.} Polarization textures in the region of external ring. (a,d) Emission of indirect excitons in the external ring region. Pattern of linear (b,e) and circular (c,f) polarization in the region of external ring. $T_{\protect bath} = 0.1$ (a-c) and 7 (d-f) K. PL intensity (g), $P_{\protect lin}$ (h), and $P_{\protect \sigma}$ (i) profiles taken along the external ring (along an axis 9 degrees of vertical positioned at maximum variation) for $T_{\protect bath} = 0.1$ (black) and 7 (red) K. PL intensity (j),  $P_{\protect lin}$ (k), and  $P_{\protect\sigma}$ (l) profiles taken across the external ring (along the x-axis positioned at maximum variation) for $T_{\protect bath} = 0.1$ (black) and 7 (red) K. The image in (j) shows the external ring at $T_{\protect bath} = 7$ K.}
\end{figure*}

The fragmentation of the external ring vanishes with increasing temperature (Fig. S7a,d,g), consistent with earlier results \cite{Butov02}. The periodic polarization textures in the region of the external ring vanish with increasing temperature (Fig. S7b,c,e,f,h,i,k,l).

Both the position of the external ring and the wavelength of the exciton density wave formed in the ring are controlled by the laser excitation indicating that the exciton density modulation in the MOES is not governed by defects in the sample. The observation of polarization textures around the MOES beads, which are not associated with defects, shows that the polarization textures do not arise due to defects.

\FloatBarrier
\subsection{The role of excitonic effects in the formation of spin textures in the exciton system}

\begin{figure*}[!htbp]
\centering
\includegraphics[width=5.5cm]{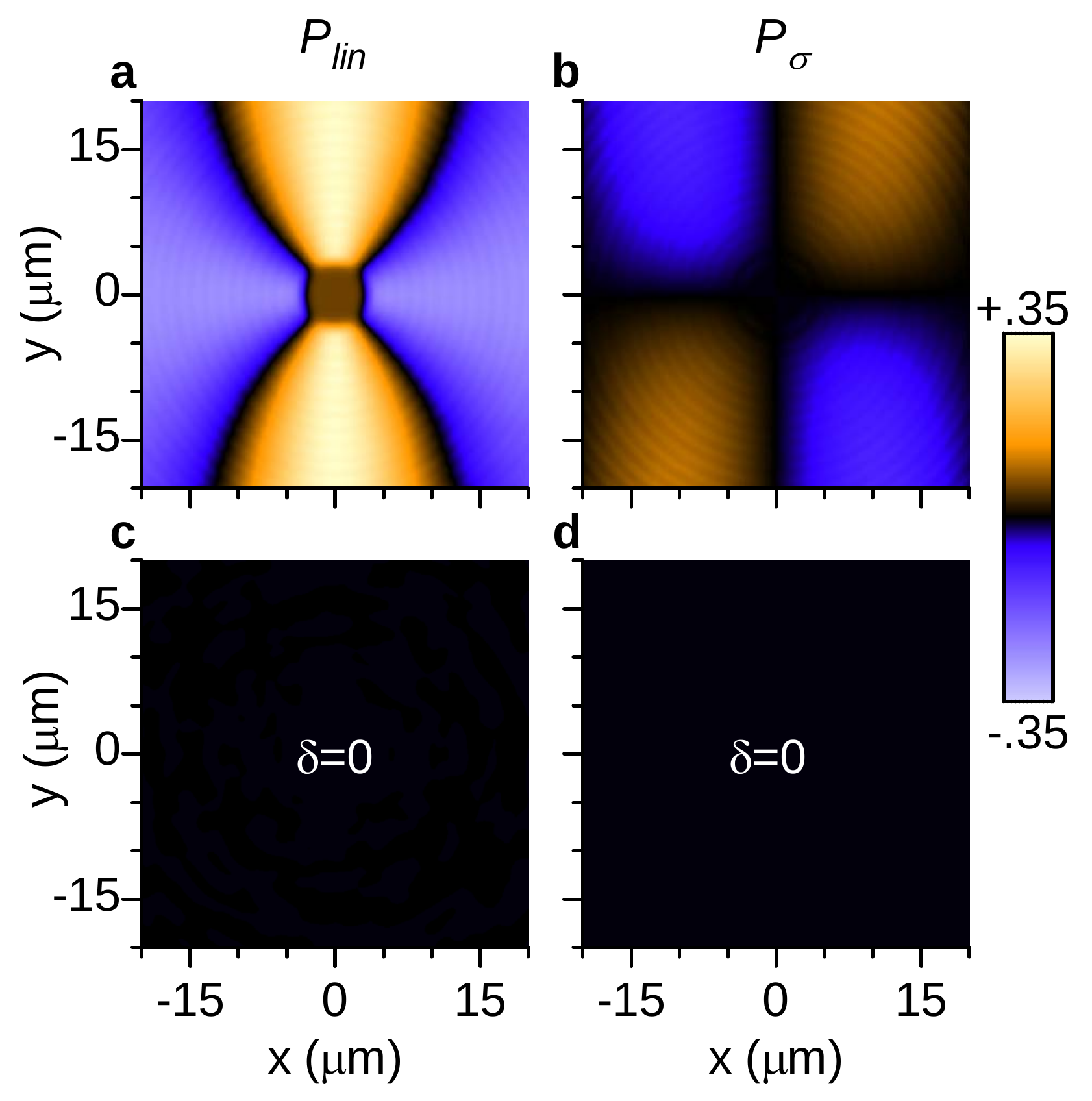}
\caption{{\bf Importance of the excitonic effects.} Simulated $P_{lin}$ and $P_\sigma$ (a,b) with the excitonic effects [nonzero excitonic terms in Hamiltonian (15), the values are presented in Section Simulations] and (c,d) without excitonic effects [zero excitonic terms in Hamiltonian (15), $\delta_b = \delta_d = E_b - E_d = 0$]. No spin texture forms in the absence of excitonic terms.}
\end{figure*}

The excitonic terms describing the spin precession in the Hamiltonian due to splitting of exciton states are present for excitons and absent for free electrons and holes. To verify the role of excitonic effects in the formation of spin textures, we performed the simulations with and without the excitonic terms in the Hamiltonian (15). No spin texture forms in the absence of excitonic terms, see Fig. S8. This shows that the excitonic effects are important for the formation of spin textures in the exciton system.

\FloatBarrier

\end{document}